# Quantum Compressed Sensing Enables Image Classification with a Single Photon

Yanshan Fan[1,2,†], Jianyong Hu[1,2,6*,†], Shuxiao Wu[1,2*], Zhixing Qiao[3], Guosheng Feng[3], Changgang Yang[1,2], Jianqiang Liu[4], Ruiyun Chen[1,2], Chengbing Qin[1,2], Guofeng Zhang[1,2], Liantuan Xiao[1,2,5,6*] and Suotang Jia[1,2]

[1]*State Key Laboratory of Quantum Optics Technologies and Devices, Institute of Laser Spectroscopy, Shanxi University, Taiyuan, 030006, China*

[2]*Collaborative Innovation Center of Extreme Optics, Shanxi University, Taiyuan, 030006, China*

[3]*College of Medical Imaging, Shanxi Medical University, Taiyuan, 030001, China*

[4]*College of Information Engineering, Shanxi Vocational University of Engineering Science and Technology, Jinzhong 030619, China*

[5]*College of Physics and Optoelectronics Engineering, Taiyuan University of Technology, Taiyuan, 030600, China*

[6]*Hefei National Laboratory, Hefei 230088, China*

[†]*These authors contributed equally.*

*\*Corresponding author E-mail address: jyhu@sxu.edu.cn; wushuxiao1@sxu.edu.cn; xlt@sxu.edu.cn*

**Abstract:** Image classification is a core task of intelligent sensing, conventionally follows a sequential "imaging-then-processing" pipeline. However, redundant high-dimensional image reconstruction is inherently inefficient, especially in photon-limited scenarios. Here we report a photon-level image classification method using quantum compressed sensing, which reformulates the classification task as a sparse-signal measurement problem directly oriented toward class labels. By exploiting the parallelism of photonic quantum superposition states, a single photon can be encoded the complete spatial information of a high-dimensional image. Through a diffractive deep neural network, we physically construct a dedicated measurement basis aligned with the class space, enabling signal-dependent adaptive compressive measurement. Ideally, our method can extract class information via a single quantum projective measurement, reducing the required number of measurements from the logarithmic scaling $O(K\log(N/K))$ of classical compressed sensing to the constant-order information-theoretic limit $M \sim K = 1$. Experimental results show that a classification accuracy of 69.0% can be achieved by using a single-photon detection event as the decision criterion, while it increases to 95.0% with four-photon detection events. This work demonstrates image classification at the energy-efficiency limit and introduces a "measurement-as-decision" framework. It provides a foundation for intelligent sensing systems that operate under extreme photon budgets and harsh environments.

**Introduction**

Image classification is one of the core tasks in modern intelligent sensing, conventionally implemented through a sequential "imaging-then-processing" architecture, that is, a high-dimensional image is first reconstructed, and low-dimensional semantic features are then extracted from it[1-4]. While this architecture functions adequately under conventional illumination, it becomes fundamentally inefficient in photon-limited scenarios, such as low-light target recognition, long-range classification, and biomedical diagnostics, where reconstructing a redundant image before making a decision wastes scarce photons and introduces unnecessary latency[5-8].

Crucially, the objective of image classification is not to recover the full image, but rather to determine its class label $c \in \{1, …, C\}$. This label can be represented as a $C$-dimensional one-hot vector, which contains exactly one nonzero entry. From the perspective of information theory[9], the classification task can be regarded as a sparse-signal decision problem with sparsity $K = 1$. It seeks to identify the position of the single nonzero component, rather than reconstructing the high-dimensional image. This insight suggests that, from first principles, the required number of measurements should scale not with the image dimension $N$, but with the sparsity $K$.

Compressed sensing (CS) was developed precisely to address the efficient acquisition of such sparse signals[10,11]. It enables the recovery of high-dimensional signals from a small number of measurements. However, classical CS performs non-adaptive measurements, in which the observation matrix is fixed independently of the signal. This fundamental constraint leads to a required measurement count of $M \propto O(K\log(N/K))$[12–14], which, for a $K = 1$ decision task, remains far from the information-theoretic lower bound of a single measurement.

Quantum mechanics offers a route to this limit by exploiting the parallelism of superposition states to enable signal-dependent adaptive measurement at the physical level[15-20]. Here we propose a quantum compressed sensing (QCS) framework that reformulates classification as a sparse-signal measurement problem directly oriented toward class labels. By leveraging the intrinsic parallelism of photonic quantum superposition states, a single photon encodes the complete spatial information of a high-dimensional image. A diffractive deep neural network (D$^2$NN) physically constructs a dedicated measurement basis aligned with the class space, realizing domain alignment between the measurement domain and the sparse label domain. This enables signal-dependent adaptive compressive measurement, allowing class information to be extracted via a single quantum projective measurement under ideal conditions. Our method reduces the required number of measurements from the logarithmic scaling $O(K\log(N/K))$ of classical CS to the constant-order information-theoretic limit $M \sim K = 1$. Experimental results show that a classification accuracy of 69.0% can be achieved by using a single-photon detection event as the decision criterion, which increases to 95.0% with four-photon detection events. This work not only demonstrates an image classification method approaching the energy-efficiency limit, but also establishes a "measurement-as-decision" information-processing paradigm. By shifting the boundary between sensing and computation, this framework lays the foundation for next-generation intelligent sensing systems operating under extreme photon budgets and harsh environments[21-26].

**QCS-based image classification model**

We define quantum compressed sensing (QCS) as a framework that leverages quantum resources, such as superposition and entanglement, to construct compressive measurements that can adapt to the signal at the physical level, thereby approaching the information-theoretic limit of sampling efficiency. This is fundamentally different from previous studies that applied classical CS to quantum systems[27-31]. In this work, we apply QCS to image classification by reformulating the task as a sparse-signal measurement problem with sparsity $K = 1$.

*Problem reformulation.* Image classification aims to assign an input image to one of $C$ mutually exclusive classes. Specifically, any input image $x \in \mathbb{R}^N$ ($N$ is the total number of image pixels) can be represented by a one-hot vector $s = e_c \in \{0,1\}^C$, where $e_c$ is the $c$-th standard basis vector and $c \in \{1, …, C\}$ is the class index. Therefore, the classification task can be equivalently expressed as the determination of the position of the only nonzero component in the one-hot class label, and is essentially a measurement-and-decision problem with sparsity $K = 1$.

Under the classical CS framework, the measurement cost is primarily determined by the reconstruction of the $N$-dimensional image, with a lower bound of $O(K\log(N/K))$. However, for a $K = 1$ class-decision task, information theory dictates that, under ideal adaptive measurement conditions, a single measurement suffices in principle to read out the class label[32,33]. The QCS-based image classification model established in this work is intended to approach this theoretical limit by exploiting quantum physical resources. This model can be summarized in four steps: quantum probe-state preparation, linear mapping from the signal to a quantum state, domain-alignment evolution, and projective measurement. Fig. 1a shows the theoretical model and physical implementation flow of this work.

***Four-step model of QCS image classification.*** If we map the sparse signal (the one-hot class vector) onto a quantum state via an image-dependent encoding, then it is translated to a quantum state discrimination problem. The key idea of our method is to align the signal measurement domain with the sparse label domain with a trainable unitary evolution. Under ideal conditions, the resulting output states corresponding to different classes become pairwise orthogonal pure states, therefore, the class label can be directly revealed by performing a projective measurement. The Holevo bound guarantees that the extractable classical information reaches $\log_2 C$ bits when the states are orthogonal and equally probable, exactly matching the requirement of a *C*-class classification task[34]. The specific four-step model of QCS image classification is given below:

***Step 1: Quantum probe-state preparation.*** To encode an *N*-dimensional image onto a single quantum state, this state must be expressible as a superposition of *N* spatial eigenstates, each corresponding to one pixel. We therefore prepare a coherent state and write it in the position basis. Let $|n\rangle$ denote the position eigenstate corresponding to the *n*-th pixel. The quantum probe state is then given by:

$$|\psi_0\rangle = \sum_{n=1}^{N} a_n |n\rangle, \quad \sum_{n=1}^{N} |a_n|^2 = 1, \tag{1}$$

where $a_n \in \mathbb{C}$ is the complex amplitude. In classical CS, the universality of measurement is ensured by a pre-fixed random observation matrix that is incoherent with the signal sparse basis, i.e., a non-adaptive measurement. For QCS, the universality of measurement can be understood as the unbiased coverage of the quantum probe state over the input signal, such as all pixels of the input image. If we directly measure the initial quantum probe state, this can be viewed as a random sampling of an *N*-dimensional input. Ideally, $|a_n|^2 = 1/N$, meaning that a measurement outcome falls on each pixel with equal probability. Repeated measurements then yield a sequence of position indices that form a random sampling matrix, which satisfies the restricted isometry property with high probability. This confirms that the initial quantum probe state is capable of carrying high-dimensional information. In practice, the prepared coherent state exhibits a Gaussian spatial distribution; even so, the resulting measurement matrix still satisfies the restricted isometry property condition with high probability (see Supplementary Note S1).

***Step 2: Linear mapping from the signal to a quantum state.*** With the initial quantum probe state prepared, we now encode the image signal $x$ onto it. The image consists of $N$ pixels, each with a reflectance value $x_n$ ($0 \leq x_n \leq 1$). For each pixel, the photon is either transmitted or blocked with a probability proportional to $x_n$. Therefore, this process can be modeled as a linear signal-dependent evolution operator $\hat{U}_x$ controlled by the signal $x$ (i.e., the image). Applying $\hat{U}_x$ to the initial probe state yields the encoded quantum state:

$$|\psi_x\rangle = \hat{U}_x |\psi_0\rangle = \sum_{n=1}^{N} \sqrt{x_n} a_n |n\rangle, \tag{2}$$

where $\hat{U}_x$ is a diagonal matrix whose diagonal elements are determined by the pixel values (i.e., reflectances) $x_n$ of the image. Since the reflectance linearly modulates the probability that the photon passes through the $n$-th path, the mapping from the image $x$ to the quantum state $|\psi_x\rangle$ is linear, which ensures that the whole process can be described by a linear equation. Consequently, any subsequent probability-based measurement will yield expectation values that are linear in $x$, forming the basis for compressive signal recovery.

***Step 3: Domain-alignment evolution.*** The encoded state $|\psi_x\rangle$ still resides in the high-dimensional pixel space. To directly extract the class label from a projective measurement, we must map this state onto a low-dimensional decision space where different classes occupy distinct, non-overlapping regions, that is, domain alignment of the measurement domain and the sparse label domain. This can be achieved by a trainable D$^2$NN[35], which physically implements a linear, programmable unitary transformation $\hat{U}_c$, and the output state can be written as:

$$|\psi_{out}\rangle = \hat{U}_c |\psi_x\rangle. \tag{3}$$

By training and optimization, for an input image belonging to class $c$, the operator $\hat{U}_c$ makes the output state $|\psi_{out}\rangle$ approximate a quantum state $|\omega_c\rangle$ that is spatially localized within a predefined region $\Omega_c$ on the detection plane, that is:

$$|\psi_{out}\rangle \approx |\omega_c\rangle \quad \text{(when the input image belongs to class } c\text{)}. \tag{4}$$

where $\{|\omega_c\rangle\}_{c=1}^{C}$ denotes the target state for class *c*. Different classes are thus mapped to mutually orthogonal spatial modes $\{\Omega_c\}_{c=1}^{C}$, ensuring that a subsequent position-basis projective measurement yields outcomes that directly indicate the class label. In the experiment, the transformation $\hat{U}_c$ is realized by loading a learned phase pattern onto a spatial light modulator (SLM), as detailed in Supplementary Note S2.

*Step 4: Projective measurement.* The output state is then subjected to a position-basis projective measurement using a single-photon avalanche diode (SPAD) array. Each pixel of the array corresponds to a measurement basis $|i\rangle$, where $i = \{1, 2, …, C\}$. According to Born's rule, the probability of detecting a photon at the *i*-th pixel is:

$$P(i) = |\langle i|\psi_{out}\rangle|^2. \tag{5}$$

By design, the class region $\Omega_c$ is mapped to a specific pixel (or a small set of pixels) on the SPAD array. Under ideal conditions, for an input of class *c*, the probability $P(i=c)$ is close to unity, while the probabilities for all other classes are negligible[36-38]. Ideally, a single photon detection event is sufficient for image classification. In practice, however, dark counts, system errors, and imperfections in the D²NN reduce the reliability of classification when a single-photon detection event is used as the decision criterion.

To enhance decision reliability, we perform multiple independent measurements for the same input. Since the mapping between the signal and the quantum state in Step 2 is linear, the expectation value $\mathbb{E}[y]$ of the measurement values (detector counts) and the one-hot class-label vector *s* satisfy $\mathbb{E}[y] = \Phi s$. Here, the equivalent observation matrix $\Phi$ is determined by the input signal and the D²NN, and its matrix element satisfies $\Phi_{i,c} \propto |\langle i|\hat{U}_c\hat{U}_x|\psi_0\rangle|^2$. After *M* independent measurements, let $y_i$ denote the cumulative counts recorded by the *i*-th detector. Then the estimate of the probability $P(i)$ is:

$$\hat{P}(i) = \frac{y_i}{M}. \tag{6}$$

When *M* is sufficiently large, $\hat{P}(i)$ converges to the true $P(i=c)$. Furthermore, in the experiment, we employ pulsed-laser illumination and time-gating techniques to suppress noise and further improve the classification accuracy by using multi-photon detection events as the decision criterion. A more detailed description of the theoretical model is provided in Supplementary Note S1.

It should be emphasized that the quantum character of our QCS scheme does not arise from the use of nonclassical light resources, such as entanglement or squeezed states. Instead, it stems from fully exploiting the parallelism inherent in the spatial-mode superposition of coherent photons, thereby achieving highly efficient sensing at the information-theoretic limit. This work is not a straightforward extension of conventional $D^2NN$ to the low-light regime. Within our QCS framework, the $D^2NN$ serves to construct a measurement basis aligned with the sparse domain, while photon detection corresponds to a projective measurement of the quantum state onto that basis. In conventional $D^2NN$-based image classification, the number of measurements is at least the class-space dimension $C$. In contrast, our approach approaches the information-theoretic limit of ideal adaptive sampling, requiring as few as $M \sim K = 1$ measurements. A detailed comparison is provided in Supplementary Note S3.

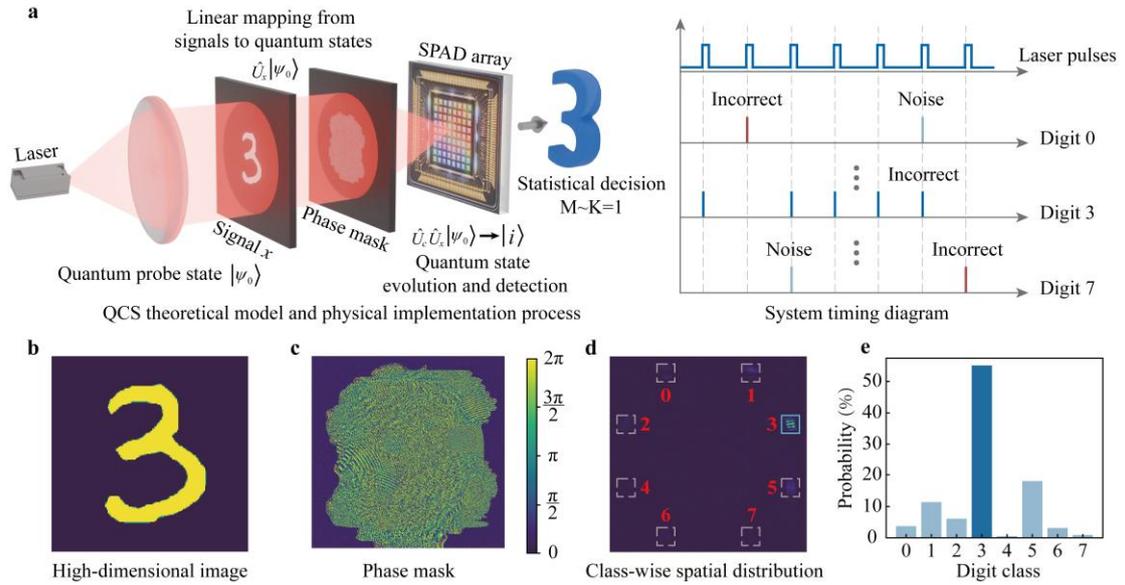

**Fig. 1 Theoretical model and experimental validation of QCS image classification. a** Schematic illustration of the QCS image classification model. The model consists of four core steps: quantum probe-state preparation, linear mapping from the signal to a quantum state (image encoding via a DMD), domain-alignment evolution (domain alignment implemented by a $D^2NN$), and projective measurement (detection using a SPAD array, followed by direct classification decision based on photon landing statistics). **b** Example of input high-dimensional image used for system validation. The digit "3" from the MNIST handwritten digit dataset is shown, which is spatially encoded onto the optical field by the DMD. **c** Phase distribution of the diffractive layer of the $D^2NN$ used for domain alignment. **d** Experimental validation of the domain-alignment. The intensity distribution on the detection plane is shown for an input digit "3", overlaid with eight predefined class regions. The optical energy is predominantly concentrated in the region corresponding to the true class, confirming that domain alignment between the signal measurement domain and the sparse label

domain has been physically realized. **e** Quantitative verification of the linear relationship between single-photon detection probability and the one-hot class-label coefficients. The measured probability distribution of photon landings across different class regions exhibits an approximately one-hot pattern.

**Experimental setup**

To verify the proposed QCS-based image classification model, we built an experimental system that strictly follows the four steps of the model. An attenuated pulsed laser serves as the initial quantum probe state and illuminates a digital micromirror device (DMD) loaded with the binarized input image. The DMD linearly modulates the propagation probability of photons along each spatial path, thereby realizing the linear mapping from the signal to a quantum state. The photons then enter a D$^2$NN implemented with a spatial light modulator (SLM), which acts as a programmable quantum processor $\hat{U}_c$ and directly maps the high-dimensional image onto a low-dimensional class space. On the detection plane, a SPAD array is used to detect the positions at which the photons collapse, with each pixel corresponding to a distinct class label. Dark counts and background noise are effectively suppressed by time-gated synchronization, and classification decisions are made based on single-photon or multi-photon detection events. Further experimental details are provided in the Methods section.

**Training of the D$^2$NN**

Domain-alignment is physically implemented by a D²NN using an SLM. We trained the phase distribution of the diffractive layer to map the encoded quantum states of different image classes onto distinct, non-overlapping regions on the detection plane. Training was performed in Python using the PyTorch framework. Free-space optical propagation was simulated with the angular spectrum method. The trainable parameters consisted of the phase values of the diffractive layer, which after training were loaded directly onto the SLM[39-43].

We designed a composite loss function to optimize the phase distribution. It contains two terms. The first term drives the output-plane intensity distribution to fit a preset target template, thereby maximizing the detection probability in the correct class region. The second term suppresses the response of confusing classes by enlarging the probability advantage between the correct class and the most confusing class, namely, the non-ground-truth class with the highest probability for the current sample. The joint action of these two terms enables the optimized diffractive layer to exhibit more pronounced class discrimination at the physical measurement stage.

The training data were derived from samples of digits 0-7 in the MNIST dataset[44], and were interpolated and binarized to adapt to the experimental system. Training was performed using the Adam optimizer to iteratively update the phase parameters until the performance on the test set converged. The final phase distribution was then directly loaded onto the SLM to realize the physical mapping from the image space to the class space. Detailed information on dataset preprocessing, network construction, and the loss function can be found in Supplementary Note S2.

**Experimental results and analysis**

We experimentally validated the proposed QCS-based image classification model. First, by visualizing the class-probability distribution, we verified that the D²NN successfully achieves domain alignment between the signal measurement domain and the sparse label domain. We then analyzed the classification performance using single-photon or multi-photon detection events as the decision criteria.

***Experimental verification of domain alignment.*** Domain alignment makes each photon detection event equivalent to a direct sampling of the class label. To verify it, we used handwritten digits 0-7 from the MNIST dataset as input signals. After being reflected by the DMD loaded with the image, the beam passed through the trained D²NN and was then detected by the SPAD array. Taking digit "3" as an example, Figs. 1b-e illustrate the domain alignment process. Fig. 1b shows the input image loaded onto the DMD, and Fig. 1c displays the phase distribution of the diffractive layer. The output intensity distribution on the detection plane is shown in Fig. 1d, where the optical energy is highly concentrated in the class region corresponding to the true class. The probability distribution of single-photon detection events over different class regions was statistically analyzed, yielding an approximately one-hot distribution, as shown in Fig. 1e. These results confirm that domain alignment between the signal measurement domain and the sparse label domain has been physically achieved.

***Classification performance.*** After verifying the domain-alignment mechanism, we further evaluated the classification performance by using single-photon and multi-photon detection events as the decision criteria. Here, a multi-photon detection event is defined as the case in which, for the same input image, $M_{pho}$ consecutive photons from multiple independent measurements all collapse into the same predefined class region

$\Omega_c$. For brevity, these two criteria are hereafter referred to as the single-photon criterion and the multi-photon criterion, with the latter specified by the photon number $M_{pho}$.

We first analyzed the photon arrival-time distribution of the optical pulses, as shown in Fig. 2a. To suppress noise while maximizing the utilization of signal photons, we set the gate width to 2.4 ns. Fig. 2b presents the core performance metric of the system, namely, the dependence of classification accuracy on the photon number $M_{pho}$ under the multi-photon criterion. Even at $M_{pho} = 1$, the average classification accuracy for the eight-class task reaches 69.0%, significantly higher than the random-guessing level (1/8 = 12.5%). As $M_{pho}$ increases, the accuracy for all classes rises rapidly and gradually saturates; for example, under the four-photon criterion, the accuracy reaches 95.0%. This behavior reflects the convergence of the QCS system from "single-shot stochastic decision" to "statistically reliable decision". Because domain alignment already concentrates the single-photon detection probability in the correct class, additional photons serve only to suppress statistical fluctuations and system noise, rather than to extract further semantic information. Consequently, the system achieves near-saturated classification performance with an extremely low photon budget.

A key trade-off is revealed in Fig. 2c, as the photon number $M_{pho}$ of multi-photon detection event increases, classification accuracy (left axis) improves because a larger $M_{pho}$ imposes a stricter consistency requirement, thereby enhancing statistical confidence. However, the occurrence probability of multi-photon detection events (right axis) decreases with $M_{pho}$. For example, although eight-photon detection events can increase the classification accuracy to 96.2%, the occurrence probability of such events is extremely low. Fig. 2d further compares two strategies: one is intensity-based decision, which makes a decision directly from the cumulative counts of all photons; the other uses multi-photon detection events as the decision criterion and triggers a high-confidence decision only when an $M_{pho}$-photon detection event occurs. The results show that multi-photon detection events yield significantly higher accuracy than intensity-based decision.

***Confusion characteristics and classification accuracy.*** To more comprehensively evaluate the system performance, we characterized the confusion structure of the classification results by means of class-conditional probability confusion matrices. Fig. 2e and f show these matrices under the single-photon criterion and the four-photon criterion, respectively. Under the single-photon criterion, off-diagonal probabilities are still observable, indicating that the measurement outputs of different classes are not yet fully separable and exhibit typical confusion patterns. These confusions arise mainly from two sources. First, morphological and structural similarities among digits lead to specific easily confused pairs. Second, limitations of training capacity, constraints of the loss function, and experimental system errors, prevent ideal domain alignment, resulting in residual probability leakage between class regions. In contrast, under the four-photon criterion, the off-diagonal probabilities are almost completely suppressed, and the confusion matrix approaches an ideal diagonal structure, reflecting the effective suppression of measurement noise under the multi-photon criterion.

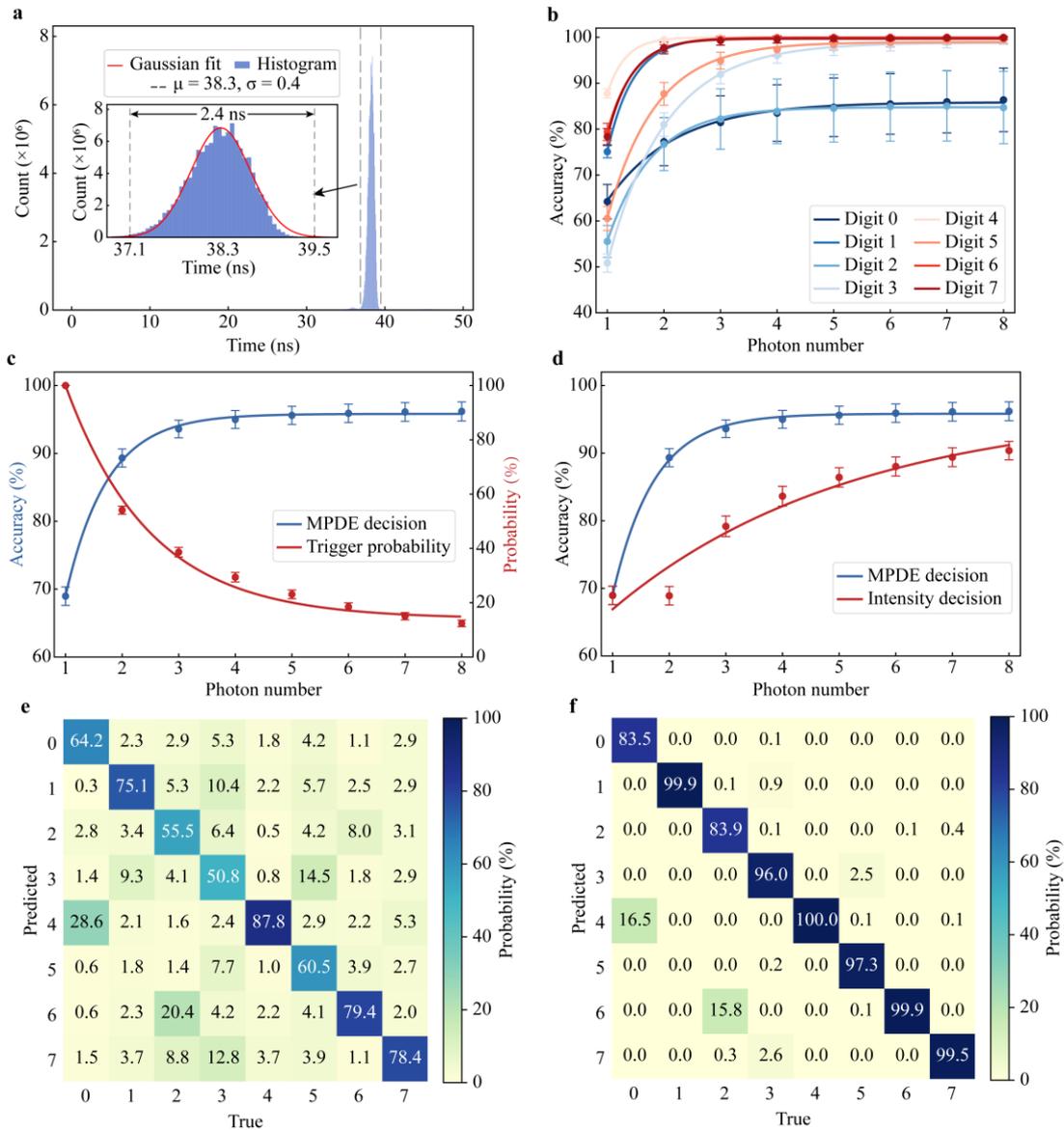

**Fig. 2 Performance characterization of the QCS-based image classification model. a** Statistical distribution of photon arrival times. The gray dashed lines indicate the time-gating window. **b** Classification accuracy as a function of the photon number $M_{pho}$ under the multi-photon criterion.

The eight curves correspond to input images of digits 0 to 7, respectively. For all classes, the accuracy increases rapidly with $M_{pho}$ at first and then gradually approaches saturation. **c** Intrinsic trade-off under the multi-photon criterion. The average classification accuracy (left axis, blue solid line) increases with $M_{pho}$, while the occurrence probability of multi-photon detection events (right axis, red solid line) decreases. **d** Performance comparison between two decision methods. The accuracy under the multi-photon criterion (blue solid line) is consistently higher than that of intensity-based decision (red solid line). **e**, **f** Class-conditional probability confusion matrices under the single-photon criterion (**e**) and the four-photon criterion (**f**). Off-diagonal elements in **e** show noticeable class confusion, whereas **f** exhibits an almost perfectly diagonal distribution. The total number of test images is 504.

**Conclusion**

In this work, we proposed and experimentally demonstrated a QCS-based image classification model that reformulates the task as a sparse-signal measurement problem with sparsity $K = 1$. By exploiting the intrinsic parallelism of photonic quantum superposition states, a single photon encodes the full spatial information of a high-dimensional image. A D²NN physically constructs a measurement basis aligned with the class space, enabling signal-dependent adaptive compressive measurement that bridges the sensing domain and the sparse label domain. Under ideal conditions, this approach extracts class information via a single quantum projective measurement, reducing the required number of measurements from the logarithmic scaling $O(K\log(N/K))$ of classical CS to the constant-order information-theoretic limit $M \sim K = 1$. Experimentally, using a single-photon detection event as the decision criterion yields a classification accuracy of 69.0%, which increases to 95.0% with four-photon detection events. These results demonstrate that the system approaches the fundamental energy-efficiency limit for optical classification. More broadly, this work establishes a "measurement-as-decision" information-processing paradigm, shifting the traditional "imaging-then-classification" architecture to a physically intelligent sensing paradigm. It lays a principled foundation for next-generation intelligent sensing systems operating under extreme photon budgets and harsh environments.

**Methods**

The QCS-based image classification model constructed in this work fully implements the four-step theoretical model of QCS. The experimental setup is shown in Fig. 3 and consists of the following four modules.

***Quantum probe-state preparation module.*** A pulsed laser (633 nm wavelength) served as the coherent light source. A variable optical attenuator attenuated the average photon number per pulse to the single-photon level. After beam expansion and collimation, the beam illuminated the DMD onto which the high-dimensional image was loaded. The quantum probe state can be represented as a superposition state over all spatial pixel paths, providing a physical carrier for the parallel encoding of high-dimensional image information and ensuring universal sensing capability for different input images.

*Linear mapping from the signal to a quantum state module.* The image to be classified was preprocessed into a binarized image and loaded onto the DMD. Each micromirror of the DMD corresponds to one image pixel, and its "on"/"off" state linearly modulates the propagation probability of an incident photon along each spatial path, evolving the quantum state that carries the image information into $|\psi_x\rangle$. This completes the linear mapping from the signal to a quantum state.

*Domain-alignment evolution module.* The photons then entered the $D^2NN$, whose diffractive layer consisted of a reflective phase SLM. The SLM was loaded with the trained phase distribution of the diffractive layer to realize the domain-alignment operator $\hat{U}_c$, thereby concentrating the output photon distribution on the detection plane onto the detector pixels corresponding to the image class labels. The training process of the $D^2NN$ is described in detail in Supplementary Note S2.

*Projective measurement module.* A 32×32 SPAD array was used as the detector on the detection plane. Eight pixels of the array were selected for quantum projective measurement, each corresponding to one measurement basis. The pulsed laser and the SPAD array were synchronously triggered, and time gating was applied to suppress dark counts and background noise.

This experimental setup shifts the computational tasks from the digital domain to the physical evolution process, fully realizing the four steps of the QCS theory and providing a platform for intelligent sensing with ultimate efficiency. Detailed information on the device models and parameters can be found in Supplementary Note S4.

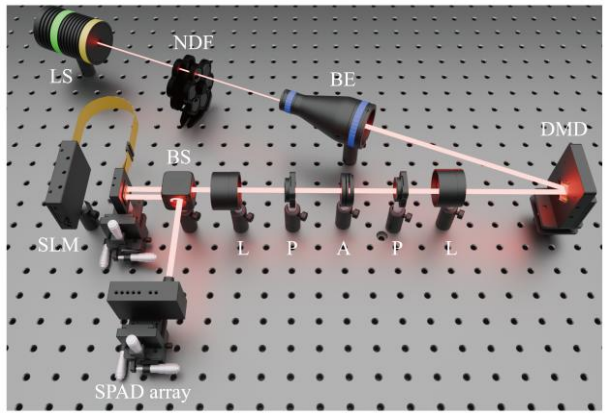

**Fig. 3 Experimental setup of the QCS-based image classification model.** LS: light source; BE: beam expander; DMD: digital micromirror device; L: lens; P: polarizer; A: aperture; BS: beam

splitter; SLM: spatial light modulator; SPAD array: single-photon avalanche diode array.

**Acknowledgements** This work was supported by Quantum Science and Technology-National Science and Technology Major Project (2021ZD0300705), Shanxi Province Basic Research Program (202503021211084), National Natural Science Foundation of China (U23A20380, 62575162, 62127817, U25D8006, U22A2091, 62305200, 62575164), National Key R&D Program of China (2022YFA1404201), Changjiang Scholars and Innovative Research Team in University of Ministry of Education of China (IRT_17R70), Overseas Expertise Introduction Project for Discipline Innovation 111 project (D18001).

**Data availability** The data that support the findings of this study are available from the corresponding author upon reasonable request. Source data are provided with this paper.

**Code availability** The code used for D²NN training and data analysis is available from the corresponding author upon reasonable request.

**Author Contributions**

The research concept and theory were developed by J.Y. H.; Experimental measurements and subsequent data analysis were carried out by Y.S. F., S.X. W. and J.Y. H.; The initial manuscript was prepared by Y.S. F. and J.Y. H.; All authors contributed to the critical revision and final approval of the manuscript.

**Competing interests**

The authors declare no competing interests.

Supplementary Information is available for this paper.

Correspondence and requests for materials should be addressed to Jianyong Hu.